\documentclass[conference]{IEEEtran}
\IEEEoverridecommandlockouts 
\def\BibTeX{{\rm B\kern-.05em{\sc i\kern-.025em b}\kern-.08em
    T\kern-.1667em\lower.7ex\hbox{E}\kern-.125emX}}
\usepackage{cite}
\usepackage{amsmath,amssymb,amsfonts}
\usepackage{algorithmic}
\usepackage{graphicx}
\usepackage{textcomp}
\usepackage{xcolor}
\usepackage{mathtools}
\usepackage{dsfont} 
\usepackage[skip=2pt,font=small]{caption}

\newtheorem{theorem}{Theorem}[section]

\newtheorem{corollary}[theorem]{Corollary}

\newtheorem{remark}[theorem]{Remark}

\usepackage{cite}

\allowdisplaybreaks
\def \OO {O}
\def \oo {\mathrm{o}}
\newcommand{\pr}{\mathbb{P}}

\newcommand{\N}{\mathbb{N}}
\newcommand{\R}{\mathbb{R}}

\newcommand{\ii}{\mathds{1}} 

\newcommand{\hh}{\mathbb{H}(n;K)} 

\newcommand{\fsquare}{\vrule height6pt width7pt depth1pt}   
\newcommand{\myendpf}{\hfill\fsquare \\[0.1in]}   

\newcommand{\nodes}{\mathcal{N}}

\newcommand{\bP}[1]{{\mathbb{P}}\left[{#1}\right]}

\begin{document}
\title{Tight Bounds for the Probability of Connectivity \\ in Random K-out Graphs }
\author{\IEEEauthorblockN{Mansi Sood and Osman Ya\u{g}an}
\IEEEauthorblockA{Department
of Electrical and Computer Engineering and CyLab, \\
Carnegie Mellon University, Pittsburgh,
PA, 15213 USA\\
msood@cmu.edu, oyagan@ece.cmu.edu}}

\maketitle

\begin{abstract}
Random K-out graphs are used in several applications including modeling by sensor networks secured by the {\em random pairwise} key predistribution scheme, and payment channel networks. The random K-out graph with $n$ nodes is constructed as follows. Each node draws an edge towards $K$ distinct nodes selected uniformly at random. The orientation of the edges is then ignored, yielding an {\em undirected} graph. An interesting property of random K-out graphs is that  
they are {\em connected}
almost surely in the limit of large $n$ for any $K \geq2$. This means that they attain the property of being connected very {\em easily}, i.e., with far fewer edges ($O(n)$)
as compared to classical random graph models including Erd\H{o}s-R\'enyi graphs ($O(n \log n)$). 
This work aims to reveal to what extent the asymptotic 
behavior of random K-out graphs being connected  easily extends to cases where the number $n$ of nodes is {\em small}. We establish upper and lower bounds on the probability of connectivity when $n$ is finite.
Our lower bounds improve significantly upon the existing results, and indicate that random K-out graphs can attain a given probability of connectivity at much smaller network sizes than previously known. 
We also show that the established upper 
and lower bounds \emph{match} order-wise; i.e., further improvement on the order of $n$ in the lower bound is not possible.
In particular, we prove that the probability of connectivity is $1-\Theta({1}/{n^{K^2-1}})$ for all $K \geq 2$. Through numerical simulations, we show that our  bounds closely mirror the empirically observed probability of connectivity.

\end{abstract}

\begin{IEEEkeywords}
 Random Graphs, Connectivity, Wireless Sensor Networks, Security
\end{IEEEkeywords}
\section{Introduction}
Random graphs constitute an important framework for analyzing the underlying structural characteristics of complex real-world networks such as communication networks, social networks and  biological networks\cite{boccaletti2006complex, goldenberg2010survey, newman2002random}. 
A class of random graphs called the random K-out graphs is one of the earliest known models of random graphs \cite{FennerFrieze1982,Bollobas}. The random K-out graph comprising $n$ nodes, denoted by $\mathbb{H}(n;K)$, is constructed as follows. Each node draws $K$ edges towards $K$ distinct nodes chosen uniformly at random from all other nodes. The orientation of the edges is then ignored, yielding an {\em undirected} graph. Due to their unique connectivity properties, random K-out graphs have received renewed interest for analyzing secure wireless sensor networks and routing in cryptocurrency networks.

\begin{figure}[t]
\centering
\includegraphics[scale=0.17]{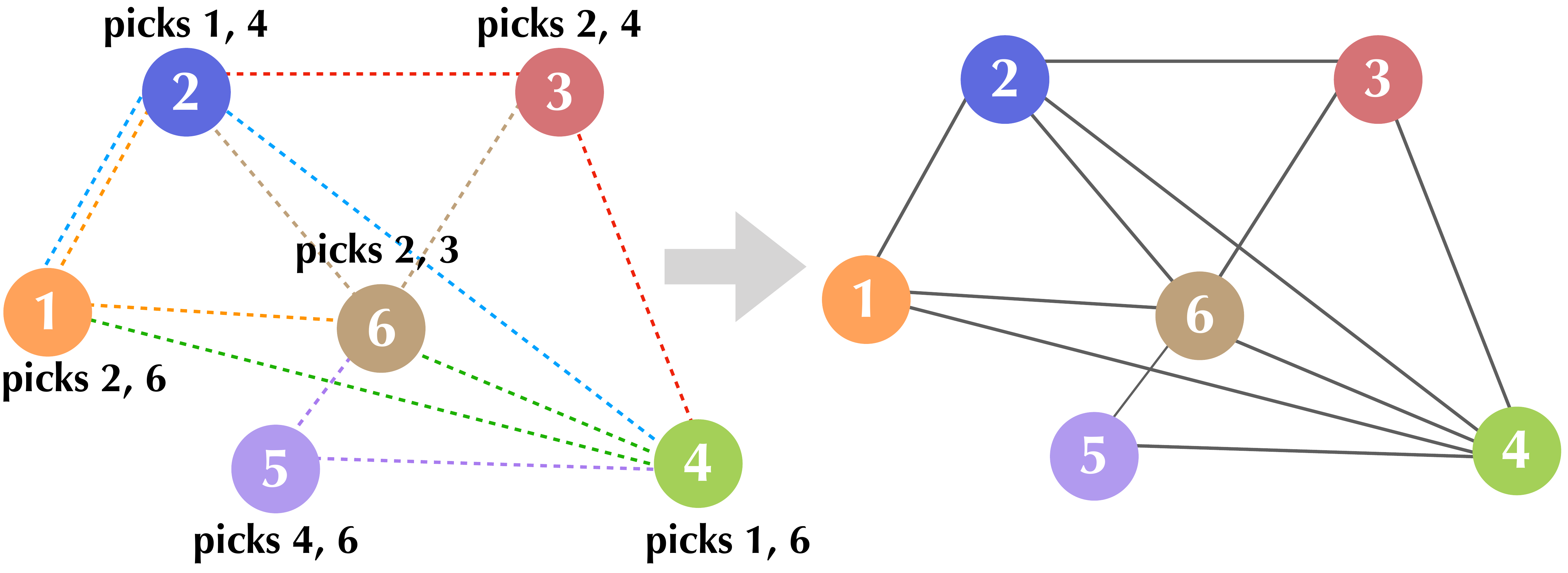}
\caption{\sl A WSN with $6$ nodes secured by the random pairwise key predistribution scheme with scheme parameter $K=2$. Each node randomly selects $K=2$ nodes and unique pairwise cryptographic keys are given to node pairs per selection. Two nodes can communicate securely if they have at least one key in common. This induces a graph with edges corresponding to node pairs that share a key.} 
\label{fig:0}
\end{figure}

In the context of wireless sensor networks (WSNs), random K-out graphs have been used extensively for evaluating strategies for secure communication.  The limited computation and communication capabilities of WSNs precludes the use of traditional key exchange protocols for establishing secure connectivity \cite{Gligor_2002,perrig2004security,XiaoSurvey}. Moreover, WSNs deployed for applications such as battlefield surveillance and environmental monitoring are vulnerable to adversarial attacks and operational failures. For facilitating secure connectivity in WSNs, Eschenauer and Gligor \cite{Gligor_2002} proposed the {\em random} predistribution of symmetric cryptographic keys. Subsequently, several variants of random key predistribution schemes have been studied; see \cite{security_survey,XiaoSurvey} and the references therein. A widely adopted approach is the random \emph{pairwise} key predistribution introduced by Chan et al.  \cite{Haowen_2003}. The random pairwise scheme is implemented in two phases. In the first phase, each sensor node is paired {\em offline} with $K$ distinct nodes chosen uniformly at random among all other sensor nodes. Next, a \emph{unique} pairwise key is inserted in the memory of each of the paired sensors. After deployment, two sensor nodes can communicate securely only if they have at least one key in common.
; see Figure~\ref{fig:0}. 
In Section~\ref{sec:Model}, we provide more details  about the implementation of this scheme. The deployment of unique, pairwise keys brings several advantages including resilience against node capture and replication attacks, and quorum-based key revocation \cite{Haowen_2003}.

In the context of cryptocurrency networks, a growing body of work is investigating the efficacy of routing protocols over different network topologies \cite{sivaraman2018high, pcn_spider2018, tang2019privacy}. A structure analogous to random K-out graphs have been proposed to make message propagation robust to {\em de-anonymization} attacks \cite[Algorithm~1]{FantiDandelion2018}. In order to make cryptocurrency networks more scalable, payment channel networks (PCNs) such as the Lightning network have been introduced. A key challenge in the design of PCNs is the trade-off between the number of edges in the network (which is constrained since each edge corresponds to funds escrowed in the PCN) and connectivity (which is desirable to facilitate  transactions between participating nodes). Given their ability to get connected with a relatively smaller number of edges, random K-out graphs offer a promising potential for informing the topological properties of such networks.
%
\par In several networked applications, {\em connectivity} is a fundamental determinant of the system performance. For instance, connectivity enables any pair of nodes to exchange messages in a communication network, or exchange funds in a cryptocurrency network. However, establishing links can be costly and often the goal is to obtain a connected network as {\em efficiently} as possible, i.e., by using the least amount of resources (links). The connectivity of random K-out graphs and their heterogeneous variants have been extensively studied \cite{FennerFrieze1982,Yagan2013Pairwise,eletrebycdc2018, mansi_globecom}. It is known \cite{Yagan2013Pairwise, FennerFrieze1982} that random K-out graphs are connected with probability tending to one (as $n \rightarrow \infty$) if and only if $K \geq 2$. 
In particular, the following zero-one law holds:
\begin{equation} 
\lim_{n \to \infty} \mathbb{P}\left[ \mathbb{H}(n;K) \text{ is connected}\right] =
\begin{cases}
1 & \mathrm{if} ~~ K\geq 2, \\
0 & \mathrm{if} ~~ K=1.
\end{cases}
\label{eq:homogeneous_zero_one_law}
\end{equation} 
A key advantage of $\hh$ is its ability to get connected very easily. With $K=2$, $\hh$ contains at most 2n edges meaning that on average, each node has a degree of less than 4. On the other hand, the classical Erd\H{o}s-R\'enyi (ER) random graph \cite{erdos61conn} requires an average degree of the order of $\log n$ for connectivity; other models with similar connectivity behavior to ER graphs include random key graphs \cite{yagan2012zero} and random geometric graphs \cite{PenroseBook}. 
Most existing results for random K-out graphs describe the behavior of the network when the number of nodes $n$ approaches $\infty$ in the form of asymptotic zero-one laws. However, in  practical scenarios, the number of nodes in the network are often constrained to be finite. This raises the need to go beyond the asymptotic results (valid for $n  \rightarrow \infty$) and obtain as tight bounds as possible for the case when $n$ is small.

Let $P(n;K)$ denote the probability of connectivity of $\hh$ as a function of the number of nodes ($n$) and the number of selections ($K$) per node. Connectivity is a monotonic increasing property in the number of edges and as a consequence $P(n;K)$ increases as $K$ increases. The case where $K=2$ corresponds to the  critical threshold  (\ref{eq:homogeneous_zero_one_law}) for connectivity. Therefore, we first focus on deriving bounds on $P(n;K)$ for the case $K=2$, and then generalize them to all $K \geq 2$. First, we derive the best known lower bound for $P(n;2)$, i.e., the probability of connectivity for $K=2$. Next, by deriving an upper bound on $P(n;2)$, we show that the lower bound \emph{matches} the upper bound order-wise, implying that further improvement on the order of $n$  is not possible. While our key focus is on the $K=2$ threshold, we also derive the lower and upper bounds for all $K \geq 2$, with our lower bound beating the existing bounds for all $K \geq 2$; see Section~\ref{sec:results} for a detailed comparison of the bounds and the empirical probability of connectivity. Moreover, to the best of our knowledge, our work is the first to derive an upper bound on $P(n;K)$, which shows that the lower bound is order-wise optimal. The \emph{matching} upper and lower bounds for the general case $K=2$ derived in this paper establish that the probability of connectivity is $1-\Theta({1}/{n^{K^2-1}})$, i.e, the probability of {\em not} being connected decays as $\Theta({1}/{n^{K^2-1}})$. Our results significantly improve the probabilistic guarantees for network designs that induce random K-out graphs.
For example, we show that $n=30$ (resp.~$n=60$) is sufficient to have a probability of connectivity of $1-10^{-4}$ (resp.~$1-10^{-5}$), while the best known previous result would indicate that $n \geq 72$
(resp.~$n \geq 150$) is necessary. 

\textbf{Organization:} In Section~\ref{sec:Model} we describe the random pairwise scheme and the resulting random K-out graphs. In Section~\ref{sec:results} we present our bounds for connectivity in random K-out graphs and compare them with existing results. We present the proofs of the lower and upper bounds, respectively in  Sections~\ref{sec:upperbound} and \ref{sec:lowerbound}, and conclude in Section~\ref{sec:conc}.

\textbf{Notation:} All limits are understood with the number of nodes $n$ going to infinity. While comparing asymptotic behavior of a pair of sequences $\{a_n\},\{b_n\}$, we use $a_n = \oo(b_n)$, $a_n=\omega(b_n)$,  $a_n = \OO(b_n)$, $a_n=\Theta(b_n)$, {\color{black}and $a_n = \Omega(b_n)$} with their meaning in the standard Landau notation. 
All random variables are defined on the same probability triple $(\Omega, {\mathcal{F}}, \mathbb{P})$.
Probabilistic statements are made with respect to this probability measure $\mathbb{P}$, and we denote the corresponding expectation operator by $\mathbb{E}$. 
The cardinality of a discrete set $A$ is denoted by $|A|$ and the set of all positive integers by $\N_0$. 

\section{Model: Random K-out Graphs}
\label{sec:Model}

The random pairwise key predistribution scheme of Chan et al. is
parametrized by two positive integers $n$ and $K$ such that $K <
n$. This scheme  is implemented as follows. Consider a network comprising of $n$ nodes {\color{black}indexed by labels} $i=1,2,\dots n$ with unique IDs: ${\rm Id}_1, \ldots , {\rm Id}_n$. Each of the $n$ nodes draws $K$ edges towards $K$ distinct nodes chosen uniformly at random from among all other nodes. Nodes $v_i$ and $v_j$ are deemed to be {\em paired} if at least one of them selected the other; i.e.,  either $v_i$ selects $v_j$, or  $v_j$ selects $v_i$, or both. Once the offline pairing process is complete, 
the set of keys to be inserted to nodes are determined as follows. For any $v_i, v_j$ that are {\em paired} with each other as described above, a unique pairwise  key $\omega_{ij}$ is generated and inserted in the memory modules of both nodes $v_i$ and $v_j$ along with the corresponding node IDs. It is important to note that 
$\omega_{ij}$ is  assigned \emph{exclusively} to nodes $v_i$ and $v_j$ to be used solely in securing the communication between them. In the post-deployment \emph{key-setup} phase, nodes first broadcast their IDs to their neighbors following which each node searches for the corresponding IDs in their key rings. Finally, nodes that have been paired  verify each others' identities through a cryptographic handshake \cite{Haowen_2003}. 

Let $\nodes:=\{1,2,\dots,n\}$ denote the set of node labels.  For each $i \in \nodes$, let $\Gamma_{n,i} \subseteq \nodes_{-i}$ denote the labels selected by node $v_i$ (uniformly at random from $\nodes_{-i}$). Specifically, for any subset $A \subseteq {\cal N}_{-i}$, we have
\begin{equation}
\pr{[\Gamma_{n,i}  = A ]} = \left \{
\begin{array}{cl}
{{n-1}\choose{K}}^{-1} & \mbox{if $|A|=K$} \\
0             & \mbox{otherwise.}
\end{array}
\right .
\label{eq:main_eqn_for_gamma}
\end{equation}
Thus, the selection of $\Gamma_{n,i}$ is done {\em uniformly}
amongst all subsets of ${\cal N}_{-i}$ which are of
size exactly $K$. Under the \emph{full-visibility} assumption,
i.e., when one-hop secure communication between a pair of sensors hinges solely on them having a common key, a WSN comprising of $n$ sensors secured by the pairwise key predistribution scheme can be modeled by a random K-out graph defined as follows. With $n=2,3, \ldots $ and
positive integer $K < n$, we say that two distinct nodes $v_i$ and $v_j$ are adjacent, denoted by $v_i \sim v_j$ if they have at least one common key in their respective key rings. More formally, 
\begin{align}
v_i \sim v_j ~~\quad \mbox{if} ~~~\quad j \in \Gamma_{n,i} \vee i \in \Gamma_{n,j}. 
\label{eq:Adjacency}
\end{align}
Let $\mathbb{H}(n;K)$ denote the undirected random graph on the
vertex set $\{ v_1, \ldots , v_n \}$ induced by the adjacency notion
(\ref{eq:Adjacency}).
In the literature on random graphs,  $\mathbb{H}(n;K)$
is often referred to as a random $K$-out graph and have been widely studied 
\cite{Bollobas,FennerFrieze1982,philips1990diameter,Yagan2013Pairwise,yavuz2017k,yagan2013scalability}.

\section{Results and Discussion}
\label{sec:results}
In this section, we present our main results,  upper and lower bounds for the probability of connectivity of $\hh$, and compare them with existing results. Throughout, we write
\[
P(n;K)
:=
\pr[{~ \mathbb{H}(n;K) ~\mbox{is connected}~}] .
\]
\subsection{Main results}
We provide our first technical result-- an upper bound for the probability of connectivity $P(n;K)$.
\begin{theorem}[Upper Bound]
\label{thm:UpperBoundForConnectivity}
For any fixed positive integer $K\geq 2$,  we have
\begin{align}
  P(n;K)
\leq  1- \frac{({K!})^K e^{-K(K+1)}}{{K+1}}\cdot \frac{1}{n^{K^2-1}}  (1+\oo(1)) 
\label{eqn:UpperBoundForConnectivity}
\end{align}{}
\end{theorem}{}
We  present the asymptotic version of the upper bound in Theorem~\ref{thm:UpperBoundForConnectivity} to make it easier to interpret; see Appendix for the more detailed bound with an explicit expression replacing the $(1+o(1))$ term in (\ref{eqn:UpperBoundForConnectivity}). The dependence of
the upper bound on the scheme parameter $K$ can be succinctly captured as follows.
\begin{remark}
For a fixed positive integer $K\geq 2$,
\begin{align}
    P(n;K)=1 - \Omega\left(\frac{1}{n^{K^2-1}}\right).
    \label{eq:upperbound-asymptotic}
\end{align}{}
\end{remark}{}
Given a fixed value of the parameter $K$ ($K \geq 2$), we derive the upper bound on the probability of connectivity by computing the likelihood of existence of isolated components comprising $K+1$ nodes. Due to space constraints, we outline the proof for the case of $K=2$ in Section~\ref{sec:upperbound} and present the full proof ($K\geq2$) in the Appendix. 
In our second main result, we derive an order-wise {\em matching} lower bound and show that the probability of connectivity is also $1-O\left(\frac{1}{n^{K^2-1}}\right)$.

\begin{theorem}[Lower Bound]
{\sl For any fixed positive integer $K\geq 2$, for all $ n \geq 4(K+2)$, we have
\begin{align}
P(n;K)
\geq 1 - c(n;K) Q(n;K)
\label{eq:LowerBoundForConnectivity}
\end{align}
where,
}
\label{thm:LowerBoundForConnectivity}
\begin{align}
c(n;K)&= \frac{e^{- (K^2-1)(1- \frac{K+1}{n})}}{\sqrt{2 \pi (K+1)}}\sqrt{\frac{n}{(n-K-1)}},
\label{eq:c(K)}\\
 Q(n;K) 
&=
\left ( \frac{K+1}{n} \right )^{K^2-1}
+ \frac{n}{2}
\left ( \frac{K+2}{n} \right )^{(K+2)(K-1)}
\label{eq:q(K)}
\end{align}
\end{theorem}
\begin{remark}
For any fixed positive integer $K\geq 2$ we have
\begin{align}
    P(n;K)
= 1 - O\left(\frac{1}{n^{K^2-1}}\right).\label{eq:lowerbound-asymptotic}
\end{align}{}
\end{remark}{}
This shows that our lower bound (\ref{eq:LowerBoundForConnectivity})  for connectivity {\em matches} our upper bound (\ref{eqn:UpperBoundForConnectivity}), and is therefore order-wise optimal. Combining (\ref{eq:upperbound-asymptotic}) and (\ref{eq:lowerbound-asymptotic}), we obtain the following result.
\begin{corollary}
{\sl For any positive integer $K\geq 2$, for all $ n \geq 4(K+2)$, we have
\begin{align}
P(n;K)
=  1 - \Theta\left(\frac{1}{n^{K^2-1}}\right)\label{eq:corollary-asymptotic}
\end{align}
}
\end{corollary}{}
The above equation 
indicates how rapidly $P(n;K)$ converges to one as $n$ grows large. 




\subsection{Previous results in \cite{Yagan2013Pairwise,FennerFrieze1982}}
We present a summary of the related lower bounds \cite{Yagan2013Pairwise, FennerFrieze1982} on the probability of connectivity. To the best of our knowledge, our work is the first to compute an upper bound on the probability of connectivity for random K-out graphs.
\subsubsection{Earlier results by Ya\u{g}an and Makowski \cite{Yagan2013Pairwise}}
It was established \cite[Theorem 1]{Yagan2013Pairwise} that for $K \geq 2$,
{
\begin{equation}
P(n;K)
\geq 1 - a(K) Q(n;K)
\label{eq:oyam_lb}
\end{equation}
holds for all $n \geq n(K)$ with $n(K) = 4(K+2)$, where
}
\begin{equation}
a(K)= e^{-\frac{1}{2} (K+1)(K-2)}.
\label{eq:a(K)}
\end{equation}




\subsubsection{Earlier results by Fenner and Frieze \cite{FennerFrieze1982}}




A lower bound for probability of connectivity can be inferred from the proof of \cite[Theorem 2.1, p. 348]{FennerFrieze1982}.
Upon inspecting Eqn. 2.2 in
\cite[p. 349]{FennerFrieze1982} with $p=0$; it can be inferred that
\begin{equation}
P(n;K) \geq 1 - b(n;K) Q(n;K)
\label{eq:BoundByFF}
\end{equation}
holds for all $n$ and $K$ such that $K < n$, where
\begin{align}
    b(n;K) = \frac{12 n}{12 n -1}\sqrt{ \frac{1}{ 2\pi (K+1) } }
\sqrt{\frac{n}{n-K-1}}.
\label{eq:b(K)}
\end{align}
Observe from
(\ref{eq:LowerBoundForConnectivity}), (\ref{eq:oyam_lb}) and (\ref{eq:BoundByFF}), that the smaller the values of $c(n;K), a(K)$ and $b(n;K)$, the better is the corresponding lower bound. As discussed in \cite{Yagan2013Pairwise}, the bound (\ref{eq:BoundByFF}) by Fenner and Frieze is tighter than (\ref{eq:oyam_lb}) when $K=2$, while (\ref{eq:oyam_lb}) is tighter than (\ref{eq:BoundByFF}) for all $K \geq 3$.
Upon examining (\ref{eq:c(K)}), (\ref{eq:a(K)}) and (\ref{eq:b(K)}), we can see that our bound given in Theorem~\ref{thm:UpperBoundForConnectivity} is tighter than both (\ref{eq:oyam_lb})  and (\ref{eq:BoundByFF}) for all $K \geq 2$. We illustrate the performance of these bounds in the succeeding discussion.

\subsection{Discussion}

\begin{figure}[!t]
\centering
\includegraphics[scale=0.37]{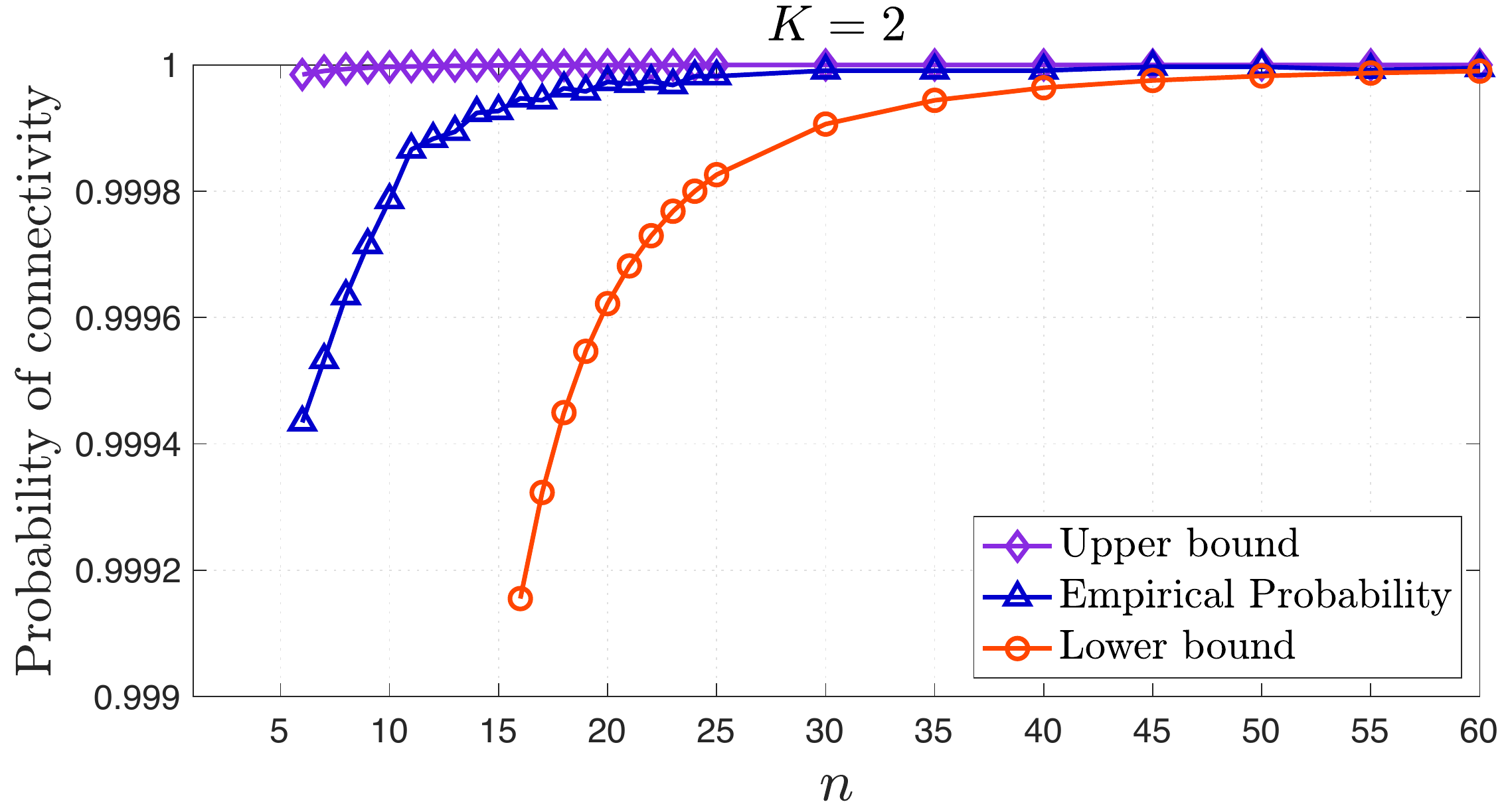}
\vspace{-1mm}
\caption{\sl A zoomed in view of our  results and empirical probability of connectivity (computed by averaging $10^6$ independent experiments for each data point) for $K=2$ as a function of $n$ for $n\geq 16$. The lower bound corresponds to Theorem \ref{thm:LowerBoundForConnectivity} and the upper bound corresponds to Theorem \ref{thm:UpperBoundForConnectivity}.
\vspace{-3mm}
}\label{fig:2}
\end{figure}

Through simulations, we study how our  upper and lower bounds compare with the empirically observed probability of connectivity. We consider a network secured by the pairwise scheme with parameter $K=2$ and compute the empirical probability of connectivity as we vary the number of nodes $n$. For each parameter pair $(n,K)$, we generate $10^6$ independent realizations of $\hh$. To obtain the empirical probability of connectivity, we divide the number of instances for which the generated graph is connected by the total number ($10^6$) of instances generated; see Figure~\ref{fig:2}. Next, we compare the lower bound for $P(n;K)$ presented in Theorem \ref{thm:LowerBoundForConnectivity} 
with the corresponding bounds in \cite{Yagan2013Pairwise,FennerFrieze1982}.  Recall from (\ref{eq:homogeneous_zero_one_law}) that $K=2$ is the critical threshold for connectivity of $\hh$ in the limit of large network size; thus, we focus on the case $K=2$ throughout the simulations. Substituting $K=2$ in (\ref{eq:q(K)}), (\ref{eq:c(K)}), (\ref{eq:a(K)}) and (\ref{eq:b(K)}), we obtain the following lower bounds on $P(n;2)$,
\begin{align}
 & \textrm{YM \cite{Yagan2013Pairwise}}:  P(n;2)  \geq 1-  \frac{155}{n^3} \label{eq:YM2}\\ 
& \textrm{FF \cite{FennerFrieze1982}}:    P(n;2) \geq 1-  \frac{155}{n^3}\cdot 
\frac{12 n/(12 n -1)}{\sqrt{ 6 \pi }}
\sqrt{\frac{n}{n-3}}
\label{eq:ff2}
\\
& \textrm{This work}  : P(n;2) \geq 1-  \frac{155}{n^3}\cdot  \frac{e^{-(3-\frac{9}{n})}}{\sqrt{6 \pi }}
    \sqrt{\frac{n}{n-3}}
    \label{eq:ourlb2}
\end{align}{}

\begin{figure}[!t]
\centering
\includegraphics[scale=0.37]{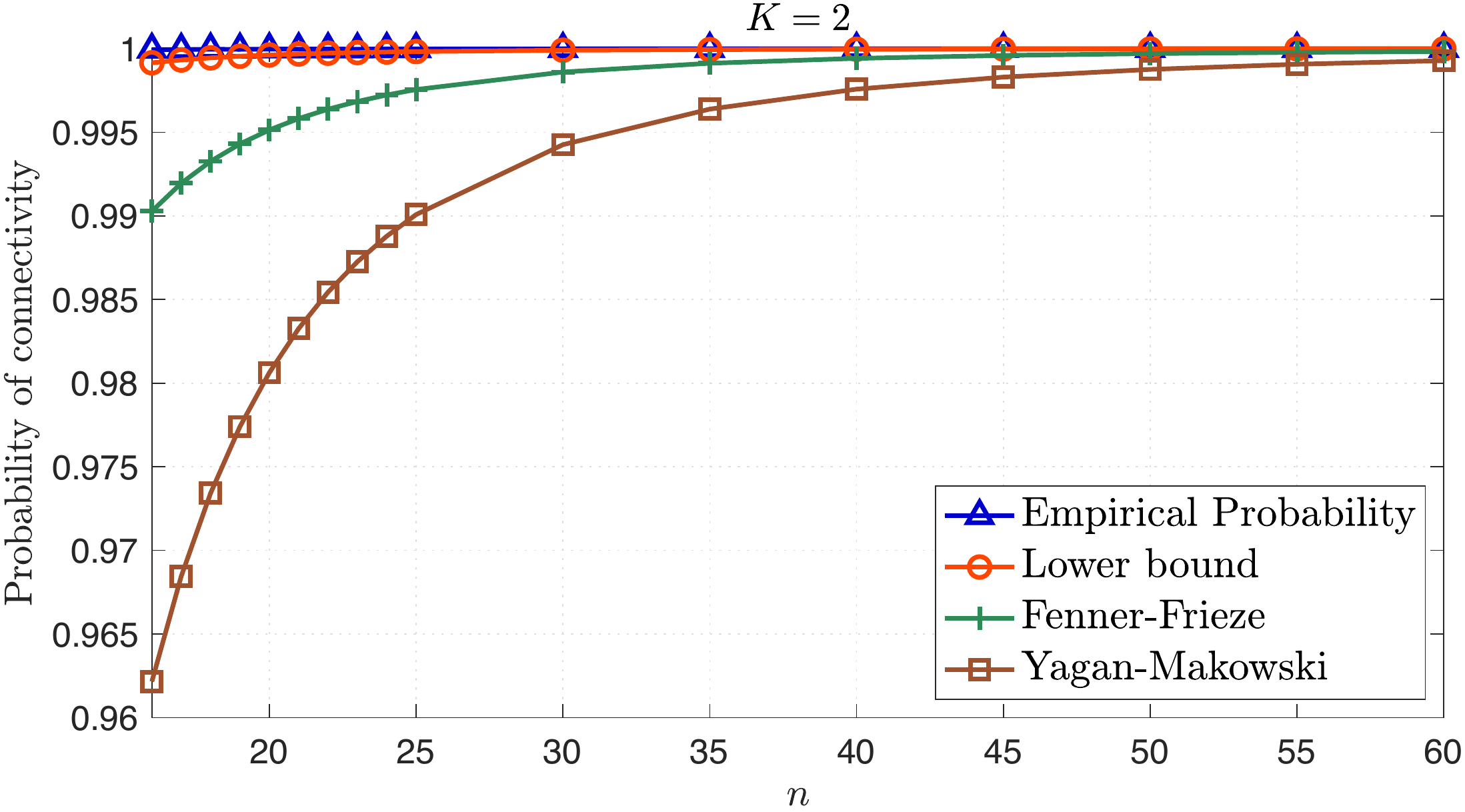}
\caption{\sl Lower bounds and empirical probability
         of connectivity (computed by averaging $10^6$ independent experiments for each data point) for $K=2$ as a function of $n$ for $n\geq 16$. Our lower bound  given in Theorem \ref{thm:LowerBoundForConnectivity} 
         significantly improves the existing lower bounds by Ya\u{g}an and Makowski \cite{Yagan2013Pairwise}, and Fenner and Frieze \cite{FennerFrieze1982}. 
         \vspace{-7mm}} \label{fig:3}
\end{figure}


With $K=2$, we plot the lower bounds (\ref{eq:LowerBoundForConnectivity}), (\ref{eq:oyam_lb}) and (\ref{eq:BoundByFF}) for comparison in Figure~\ref{fig:3}. In Table~\ref{tab:tab1}, we compare the mean number of realizations of $\hh$ generated until one disconnected realization is observed corresponding to the lower bounds (\ref{eq:YM2}), (\ref{eq:ff2}) and (\ref{eq:ourlb2}) for $K=2$.

Our results show that $\hh$ gets connected with probabilistic guarantees as high as $99.92\%$ even when $K=2$ and network consisting of as few as $16$ nodes. These results complete and complement the existing asymptotic zero-one laws for random K-out graphs.

\begin{table}[!h]
\begin{center}
\begin{tabular}{|c|c|c|c|}
   \hline
   & \multicolumn{3}{c|}{Mean number of disconnected realizations}
   \\
   \hline
$n$& Theorem~\ref{thm:LowerBoundForConnectivity}& YM \cite{Yagan2013Pairwise} & FF \cite{FennerFrieze1982}\\
   \hline
     16&1 in {\bf1183} &1 in {\bf26} & 1 in {\bf102} \\
     \hline
     20&1 in {\bf2645} &  1 in {\bf51} & 1 in {\bf 205}\\
     \hline
          25&1 in {\bf 5753} & 1 in {\bf100}&1 in {\bf409} \\
           \hline
     35&1 in {\bf 17834 } & 1 in {\bf 276} & 1 in {\bf 1145 }\\
     \hline
\end{tabular}{}
\end{center}
     \caption{\sl
          Comparison of the lower bound (\ref{eq:LowerBoundForConnectivity}) with existing lower bounds (\ref{eq:oyam_lb}) and (\ref{eq:BoundByFF}) from \cite{Yagan2013Pairwise} and \cite{FennerFrieze1982}, respectively for $K=2$. The entries in the table corresponds to the mean number of realizations of $\hh$ generated until one disconnected realization is observed.
\vspace{-4mm}}
     \label{tab:tab1}
     \end{table}

\section{Upper bound on probability of connectivity}
\label{sec:upperbound}
For easier exposition, we give a proof 
of Theorem~\ref{thm:UpperBoundForConnectivity} here
for $K=2$. 
Due to space constraints, the general version of our proof for $K\geq 2$ is given in the Appendix. For $K=2$, each node selects at least two other nodes and there can be no isolated nodes or node pairs in $\hh$. Thus, for $K=2$, the smallest possible isolated component is a {\em triangle},  i.e., a complete sub-networks over {\em three} nodes such that each node selects the other two nodes.  To derive the upper bound on connectivity, we first derive a lower bound on the probability of existence of isolated triangles in $\hh$. In the proof for the general case $(K \geq 2)$ presented in the Appendix, we investigate the existence of isolated components of size $K+1$.

Let $\Delta_{ijk}$ denote the event that nodes $v_i,v_j$ and $v_k$ form an isolated triangle in $\hh$. The number of isolated triangles in $\hh$, denoted by $Z_n$ is given by
\begin{align}
    Z_n = \sum_{1\leq i<j<k \leq n} \ii\{ \Delta_{ijk}\}
\end{align}{}Note that the existence of one or more isolated triangles  ($Z_n \geq 1$), implies that $\hh$ is \emph{not} connected. Thus, we can upper bound the probability of connectivity of $\hh$ as
\begin{align}
    &\pr [~\hh \text{~is connected}~]\nonumber \\
    &=1- \pr [~\hh \text{~is \emph{not} connected}~]\nonumber \\
   &=1-\pr[~\exists \text{~at least one isolated \emph{sub-network} in~} \hh ~]\nonumber \\
    &\leq 1-\pr[~ \exists \text{~at least one {isolated} \emph{triangle} in~} \hh ~] \nonumber \\
     &= 1-\pr[~ Z_n \geq 1 ~].
    \label{eq:p_con_upper_bound}
\end{align}{}
where,
\begin{align}
\vspace{-1mm}
[Z_n \geq 1] = \cup_{1\leq i<j<k \leq n} \ii\{\Delta_{ijk}\}.
\end{align}{}

In the succeeding discussion, we assume $K=2$ and
 use the Bonferroni inequality \cite{bonferroni} to lower bound the union of the events $\ii\{\Delta_{ijk}\}$, where $1\leq i<j<k \leq n$.
\begin{align}
&\pr[~ Z_n \geq 1 ~] \nonumber\\
& \geq \sum_{ i<j<k } \pr [~\Delta_{ijk}~]-\sum_{ i<j<k }\sum_{ x<y<z } \pr [~\Delta_{ijk}\cap\Delta_{xyz}~]
\label{eq:2out-union-lb}
\end{align}{}For all $1\leq i<j<k\leq n$ and $1 \leq x<y<z \leq n$, we have
\begin{align}
    \pr [~\Delta_{ijk}~]= \left(\frac{{1}}{{n-1 \choose 2}}\right)^3\left(\frac{{n-4 \choose 2}}{{n-1 \choose 2}}\right)^{n-3}
    \label{eq:2out-singlesum}
\end{align}{}
Moreover, note that if the sets $\{i,j,k\}$ and $\{x,y,z\}$ have one or more nodes in common, then these sets cannot simultaneously constitute isolated triangles; i.e., the events $\Delta_{ijk}$, $\Delta_{xyz}$ are mutually exclusive if $\{i, j, k\} \cap \{x, y, z\} \neq \emptyset $.
Thus, 
\begin{align}
\pr [~\Delta_{ijk} \cap\Delta_{x y z}~]=
\begin{cases}
& 0  \text{~~~if~} \{i,j,k\}\cap\{x,y,z\} \neq \phi, \\
& \left(\frac{{1}}{{n-1 \choose 2}}\right)^6\left(\frac{{n-7 \choose 2}}{{n-1 \choose 2}}\right)^{n-6} \text{otherwise}.
\label{eq:2out-doublesum}
\end{cases}
\end{align}{}We now calculate the term appearing in (\ref{eq:2out-union-lb}) in turn. We have
\begin{align}
    \sum_{ i<j<k } \pr [~\Delta_{ijk}~]
    &= {n \choose 3} \pr [~\Delta_{ijk}~]\nonumber\\
    &= {n \choose 3}\left(\frac{{1}}{{n-1 \choose 2}}\right)^3\left(\frac{{n-4 \choose 2}}{{n-1 \choose 2}}\right)^{n-3}\nonumber\\
   &= \frac{{4n}}{{3(n-1)^2 (n-2)^2}}
   \prod_{\ell=1}^2\left( 1-\frac{3}{n-\ell}\right)^{n-3} 
   \nonumber\\
   &\geq \frac{{4}}{{3n^3}}\left( 1-\frac{3}{n-2}\right)^{2n-6}
   \label{eq:2out-bound-singlesum},
\end{align}{}
and
\begin{align}
    &\sum_{ i<j<k }\sum_{ x<y<z } \pr [~\Delta_{ijk}\cap\Delta_{xyz}~] \nonumber\\
    &= {n \choose 3}{n-3 \choose 3}\left(\frac{{1}}{{n-1 \choose 2}}\right)^6\left(\frac{{n-7 \choose 2}}{{n-1 \choose 2}}\right)^{n-6}\nonumber\\
     &= \frac{{16n(n-3)(n-4)(n-5)}}{{9 (n-1)^5(n-2)^5}}
     \prod_{\ell=1}^2\left( 1-\frac{6}{n-\ell}\right)^{n-6} 
    \nonumber\\
     &\leq \frac{{16n^4}}{{9 (n-2)^{10}}}\left( 1-\frac{6}{n-1}\right)^{2n-12}
      \label{eq:2out-bound-doublesum}.
\end{align}{}
Substituting (\ref{eq:2out-bound-singlesum}) and (\ref{eq:2out-bound-doublesum}) in (\ref{eq:2out-union-lb}), we obtain
\begin{align}
&\pr[~ Z_n \geq 1 ~] \nonumber\\
& \geq \sum_{ i<j<k } \pr [~\Delta_{ijk}~]-\sum_{ i<j<k }\sum_{ x<y<z } \pr [~\Delta_{ijk}\cap\Delta_{xyz}~]\nonumber\\
& \geq \frac{{4}}{3{n^3}}\hspace{-.5mm}\left( \hspace{-.5mm}1\hspace{-.5mm}-\hspace{-.5mm}\frac{3}{n-2}\hspace{-.5mm}\right)^{2n\hspace{-.5mm}-6} \hspace{-1.5mm}-\frac{{16n^4}}{{9 (n-2)^{10}}}\left(\hspace{-.5mm} 1\hspace{-.5mm}-\hspace{-.5mm}\frac{6}{n-1}\hspace{-.5mm}\right)^{2n-12}\nonumber\\
& =\frac{{4 e^{-6}}}{3{n^3}}(1+\oo(1)) \label{eq:upperrbound}
\end{align}{}
Reporting this into (\ref{eq:p_con_upper_bound})
leads to establishing
Theorem \ref{thm:UpperBoundForConnectivity}
for $K=2$.
More compactly, this result can be stated
as $P(n;2)=1-\Omega\left(\frac{1}{n^3}\right)$.
We prove the more general result for $K \geq 2$ 
in the Appendix. 
The next Section is devoted establishing a {\em matching} lower bound on the probability of connectivity.

\section{Lower bound on probability of connectivity}
Fix $n=2,3, \ldots $ and consider a fixed positive integer $K$.
The conditions
\begin{equation}
2 \leq K  \quad \mbox{and} \quad e(K+2) < n
\label{eq:OneLawConditions}
\end{equation}
are enforced throughout. Note that the condition $e(K+2) < n$ automatically implies $K<n$.

\label{sec:lowerbound}
\subsection{Preliminaries}
\noindent Before proceeding with the proof, we discuss one of the key steps which distinguishes our proof and improves upon existing\cite{Yagan2013Pairwise,FennerFrieze1982} bounds.  In contrast to the standard bound ${n \choose r} \leq \left(\frac{ne}{r} \right)^r$ used in \cite{Yagan2013Pairwise}, we upper bound ${n \choose r}$ using a variant \cite{stirlingremark} of Stirling formula. For all $x=1,2,\dots$, we have
\begin{align}
    \sqrt{2 \pi} x^{x+0.5} e^{-x} e^{\frac{1}{12x+1}} < x! <    \sqrt{2 \pi} x^{x+0.5} e^{-x} e^{\frac{1}{12x}}, \label{eq:stirling}
\end{align}{}
which gives
\begin{align}
{n \choose{ r}}& \leq \frac{1}{\sqrt{2 \pi}} \left(\frac{n}{n-r}\right)^{n-r} \left( \frac{n}{r}\right)^r \frac{\sqrt{n}}{\sqrt{n-r}\sqrt{r}} \nonumber\\
&\qquad \cdot \exp\left\{ \frac{1}{12n}-\frac{1}{12(n-r)+1}-\frac{1}{12r+1}\right\}\nonumber\\
& \leq \frac{1}{\sqrt{2 \pi}} \left(\frac{n}{n-r}\right)^{n-r} \left( \frac{n}{r}\right)^r \frac{\sqrt{n}}{\sqrt{n-r}\sqrt{r}},
\label{eq:stirlingtakeaway}
\end{align}{}
since $$\frac{1}{12n}-\frac{1}{12(n-r)+1}-\frac{1}{12r+1} < 0$$
Using the upper bound for ${n \choose r}$ as presented in (\ref{eq:stirlingtakeaway}) eventually leads to the factor 
$e^{- (K^2-1)(1- \frac{K+1}{n})}$ improvement in the lower bound on probability of connectivity in Theorem~\ref{thm:LowerBoundForConnectivity}. Next, we note that for $0\leq K \leq x \leq y$, 
\begin{align}
 \frac{ {x \choose K} }{ {y \choose K} }
= \prod_{\ell=0}^{K-1}
\left (
\frac{x-\ell}{y-\ell} \right ) \leq \left ( \frac{x}{y}
\right )^K \label{eq:ratio}   
\end{align}{}
since $\frac{x-\ell}{y-\ell}$ decreases as $\ell$ increases from
$\ell = 0$ to $\ell=K-1$.
Lastly, for all $x \in \R$, we have
\begin{align}
    1  \pm x  &\leq e^{\pm x}. \label{eq:kcon_1pmx}
\end{align}

\subsection{Proof of Theorem~\ref{thm:LowerBoundForConnectivity}}
If $\mathbb{H} (n;K)$ is {\em not} connected, then there exists a non-empty subset $S$ of nodes that is isolated. Further, since each node is paired with at least $K$ neighbors, $|S|\geq K+1$. Let $C_n (K)$ denote the event that $\mathbb{H}(n;K)$ is
connected. We have 
\begin{equation}
C_n(K)^c
\subseteq
\bigcup_{S \in \mathcal{P}_n: ~ |S| \geq K+1} ~ B_n (K ; S)
\label{eq:BasicIdea}
\end{equation}
where $\mathcal{P}_n$ stands for the collection of all non-empty
subsets of ${\cal N}$. Let $\mathcal{P}_{n,r} $ denotes the collection of all subsets
of ${\cal N}$ with exactly $r$ elements.
A standard union bound argument yields
\begin{eqnarray}
\bP{ C_n(K)^c }
&\leq &
\sum_{ S \in \mathcal{P}_n: K+1 \leq |S| \leq \lfloor \frac{n}{2} \rfloor }
\bP{ B_n (K ; S) }
\nonumber \\
&=&
\sum_{r=K+1}^{ \lfloor \frac{n}{2} \rfloor }
\left ( \sum_{S \in \mathcal{P}_{n,r} } \bP{ B_n (K; S) } \right ).
\label{eq:BasicIdea+UnionBound}
\end{eqnarray}
For each $r=1, \ldots , n$, let
$B_{n,r} (K) = B_n (K ; \{ 1, \ldots , r \} )$. Under the enforced
assumptions, exchangeability implies
\[
\bP{ B_n (K ; S) } = \bP{ B_{n,r} (K ) }, \quad S \in
\mathcal{P}_{n,r}
\]
and since $|\mathcal{P}_{n,r} | = {n \choose r}$, we have
\begin{equation}
\sum_{S \in \mathcal{P}_{n,r} } \bP{ B_n (K ; S) } 
= {n \choose r} ~ \bP{ B_{n,r}(K ) } 
\label{eq:ForEach=r}
\end{equation}
Substituting (\ref{eq:ForEach=r})
into (\ref{eq:BasicIdea+UnionBound}) we obtain 
\begin{align}
\bP{ C_n(K)^c }& \leq \sum_{r=K+1}^{ \lfloor \frac{n}{2} \rfloor }{n \choose r} ~ \bP{ B_{n,r}(K ) } .
\nonumber\\
&\leq 
\sum_{r=K+1}^{ \lfloor \frac{n}{2} \rfloor } {n \choose r}\hspace{-1mm}
\left({ {r-1}\choose K }\over { {n-1} \choose K } \right)^{\hspace{-.5mm}r}\hspace{-1mm}
\left(\hspace{-1mm}{ {n-r-1}\choose K }\over { {n-1} \choose K } \hspace{-1mm}\right)^{n-r}.\label{eq:BasicIdea+UnionBound2}
\end{align}



 Using (\ref{eq:ratio}) in
(\ref{eq:BasicIdea+UnionBound2}) together with  (\ref{eq:stirlingtakeaway}), 
we conclude that
\begin{align}
& \bP{ C_n(K)^c }
\nonumber \\
& \leq \sum_{r=K+1}^{ \lfloor \frac{n}{2} \rfloor } {n \choose r} \left(\frac{r-1}{n-1}\right)^{rK}
\left(1-\frac{r}{n-1}\right)^{(n-r)K}\\
& \leq \sum_{r=K+1}^{ \lfloor \frac{n}{2} \rfloor } {n \choose r} \left(\frac{r}{n}\right)^{rK}
\left(1-\frac{r}{n}\right)^{(n-r)K}\nonumber\\
& \leq \sum_{r=K+1}^{ \lfloor \frac{n}{2} \rfloor }
\frac{1}{\sqrt{2 \pi}} \left(\frac{n}{n-r}\right)^{n-r} \left( \frac{n}{r}\right)^r \frac{\sqrt{n}}{\sqrt{n-r}\sqrt{r}}\nonumber\\ &\qquad \cdot \left(\frac{r-1}{n-1}\right)^{rK}
\left(1-\frac{r}{n-1}\right)^{(n-r)K}
\nonumber \\
& =
\sum_{r=K+1}^{ \lfloor \frac{n}{2} \rfloor }\frac{\sqrt{n}}{\sqrt{2 \pi}\sqrt{n-r}\sqrt{r}}
\left(\frac{r}{n}\right)^{r(K-1)}
\left(1-\frac{r}{n}\right)^{(n-r)(K-1)}
\nonumber \\
&\leq
\hspace{-1mm}\sum_{\hspace{-.5mm}r=K+1}^{ \lfloor \frac{n}{2} \rfloor }\hspace{-.5mm}\frac{\sqrt{n}}{\sqrt{2 \pi}\sqrt{n-r}\sqrt{r}}
\left(\hspace{-.5mm}\frac{r}{n}\hspace{-.5mm}\right)^{r(K-1)}
\hspace{-1.5mm}e^{-\left(\frac{r}{n}\right){(n-r)(K-1)}}
\label{eq:eapply},
\end{align}
where (\ref{eq:eapply}) follows from (\ref{eq:kcon_1pmx}). For $K+1 \leq r \leq  \lfloor \frac{n}{2} \rfloor,$ we have
\begin{align}
r(n-r) \geq (K+1)(n-K-1) 
\end{align}
Substituting in (\ref{eq:eapply}),
\begin{align}
&{\bP{ C_n(K)^c }}\nonumber \\
 & \leq \sum_{r=K+1}^{ \lfloor \frac{n}{2} \rfloor }\frac{\sqrt{n}}{\sqrt{2 \pi}\sqrt{n-K-1}\sqrt{K+1}}
\left(\frac{r}{n}\right)^{r(K-1)}\nonumber\\ &\qquad \cdot
e^{-\left(\frac{K+1}{n}\right){(n-K-1)(K-1)}}
\nonumber \\
& =
\sum_{r=K+1}^{ \lfloor \frac{n}{2} \rfloor }
\left(\frac{r}{n}\right)^{r(K-1)} 
\frac{e^{- (K^2-1)(1- \frac{K+1}{n})}}{\sqrt{2 \pi (K+1)}}\sqrt{\frac{n}{(n-K-1)}}.
\nonumber  \\
& = c(n;K)
\sum_{r=K+1}^{ \lfloor \frac{n}{2} \rfloor }
\left(\frac{r}{n}\right)^{r(K-1)}
\nonumber \\
& = c(n;K)\hspace{-1mm}
\left(\hspace{-0.5mm}\frac{K+1}{n}\hspace{-0.5mm}\right)^{K^2-1}
\hspace{-1.5mm} +
c(n;K)\hspace{-1.5mm}
\sum_{r=K+2}^{ \lfloor \frac{n}{2} \rfloor }
\left(\frac{r}{n}\right)^{r(K-1)}
\label{eq:con_1_law_last_step} 
\end{align}
with $c(n;K)$ given by (\ref{eq:c(K)}). Due to space constraints we in present the  sequence of steps leading to the final bound in Theorem~\ref{thm:LowerBoundForConnectivity} in the Appendix.

\section{Conclusions}
In this work we derive upper and lower bounds for connectivity for random K-out graphs when the number of nodes is finite. Our {matching} upper and lower bounds prove that the probability of connectivity is $1-\Theta({1}/{n^{K^2-1}})$ for all $K \geq 2$. 
Our lower bound is shown to significantly improve the existing ones. In particular, our results further strengthen the applicability of 
random K-out graphs as an efficient way to construct a connected network topology even when the number of nodes is {\em small}.
 It would be interesting to pursue further applications of K-out graphs in the context of cryptographic payment channel networks.
\label{sec:conc}



\section*{Acknowledgements}
This work has been supported in part by the
National Science Foundation through grant CCF \#1617934.
\bibliographystyle{IEEEtran}
\bibliography{IEEEabrv,references}
\newpage
\section*{Appendix}

\subsection{Upper bound on probability of connectivity for $K\geq2$}
\label{sec:genupperbound}
In Section~\ref{sec:upperbound} we proved Theorem~\ref{thm:UpperBoundForConnectivity} for the case $K=2$. In this section, we prove the upperbound for the general case $K \geq 2$. Let $K$ be a fixed positive integer such that $ K \geq 2$. Let $\Delta_{i_1 \dots i_{K+1}}$ denote the event that nodes $v_i,v_j,\dots,v_{K+1}$ form an isolated component in $\hh$. The number of such isolated components of size $K+1$ in $\hh$, denoted by $Z_n$ is given by
\begin{align}
    Z_n = \sum_{1\leq i_1<i_2 \dots < i_{K+1} \leq n} \ii\{ \Delta_{i_1 \dots i_{K+1}}\}
\end{align}{}
Note that the existence of one or more isolated components of size $K+1$  ($Z_n \geq 1$), implies that $\hh$ is \emph{not} connected. We can upper bound the probability of connectivity of $\hh$ as
\begin{align}
    &\pr [~\hh \text{~is connected}~]\nonumber \\
    &=1- \pr [~\hh \text{~is \emph{not} connected}~]\nonumber \\
   &=1-\pr[~\exists \text{~at least one isolated \emph{sub-network}}  ~]\nonumber \\
    &\leq 1-\pr[~ \exists \text{~at least one {isolated} component of size $K+1$}  ~] \nonumber \\
     &= 1-\pr[~ Z_n \geq 1 ~].
    \label{eq:gen_p_con_upper_bound}
\end{align}{}
where,
\begin{align}
\{Z_n \geq 1\} = \bigcup_{1\leq i_1<i_2 \dots < i_{K+1} \leq n} \ii\{\Delta_{i_1 \dots i_{K+1}}\}.
\end{align}{}

In the succeeding discussion, we use the Bonferroni inequality \cite{bonferroni} to lower bound the union of the events given in (\ref{eq:gen_p_con_upper_bound}).
\begin{align}
&\pr[~ Z_n \geq 1 ~] \nonumber\\
& \geq \sum_{ i_1<i_2 \dots < i_{K+1} } \pr [~\Delta_{i_1 \dots i_{K+1}}~]\\
&\qquad -\sum_{i_1<i_2 \dots < i_{K+1} }\sum_{ j_1<j_2 \dots < j_{K+1}} \pr [~\Delta_{i_1 \dots i_{K+1}}\cap\Delta_{j_1 \dots j_{K+1}}~]
\label{eq:gen_2out-union-lb}
\end{align}{}
For all $1\leq i_1<i_2 \dots < i_{K+1}\leq n$ and $1 \leq j_1<j_2 \dots < j_{K+1}\leq n$, we have
\begin{align}
    \pr [~\Delta_{i_1 \dots i_{K+1}}~]= \left(\frac{{1}}{{n-1 \choose K}}\right)^{K+1}\left(\frac{{n-K-2 \choose K}}{{n-1 \choose K}}\right)^{n-K-1}
    \label{eq:gen_2out-singlesum}
\end{align}{}
Moreover, note that if the sets $\{i_1,\dots,i_{K+1}\}$ and $\{j_1,\dots,j_{K+1}\}$ have one or more nodes in common, then these sets cannot simultaneously constitute isolated components.
Thus, $\pr [~\Delta_{i_1 \dots i_{K+1}} \cap\Delta_{j_1 \dots j_{K+1}}~]=$
\begin{align}
\begin{cases}
& 0  \text{~~~if~~~} \{i_1,\dots,i_{K+1}\}\cap\{j_1,\dots,j_{K+1}\} \neq \phi, \\
& \left(\frac{{1}}{{n-1 \choose K}}\right)^{2(K+1)}\left(\frac{{n-2K-3 \choose K}}{{n-1 \choose K}}\right)^{n-2(K+1)} \text{otherwise}. \label{eq:gen_2out-doublesum}
\end{cases}
\end{align}{}
We now calculate the term appearing in (\ref{eq:gen_2out-union-lb}) in turn. We have
\begin{align}
    &\sum_{ i_1<i_2 \dots < i_{K+1}} \pr [~\Delta_{i_1\dots i_{K+1}}~]\nonumber\\
    & = {n \choose {K+1}} \pr [~\Delta_{i_1 \dots i_{K+1}}~]\nonumber\\
    & = {n \choose {K+1}}\left(\frac{{1}}{{n-1 \choose K}}\right)^{K+1}\left(\frac{{n-K-2 \choose K}}{{n-1 \choose K}}\right)^{n-K-1}\nonumber\\
       & = \frac{({K!})^K n}{{K+1}}\cdot \left(\frac{(n-K-1)!}{(n-1)!}\right)^{K} \prod_{\ell=1}^{K}\cdot\left(1-\frac{K+1}{n-\ell}\right)^{n-K-1} \nonumber\\
   & \geq  \frac{({K!})^K}{{K+1}}\cdot \frac{1}{n^{(K^2-1)}} \prod_{\ell=1}^{K}\cdot\left(1-\frac{K+1}{n-\ell}\right)^{n-K-1} \nonumber\\
      & \geq  \frac{({K!})^K}{{K+1}}\cdot \frac{1}{n^{(K^2-1)}} \cdot\left(1-\frac{K+1}{n-K}\right)^{K(n-K-1)} \label{eq:gen_2out-bound-singlesum}\\
       & = \frac{({K!})^K e^{-K(K+1)}}{{K+1}}\cdot \frac{1}{n^{(K^2-1)}}  (1+\oo(1)) \nonumber
\end{align}{}
where (\ref{eq:gen_2out-bound-singlesum}) is plain from the observation that for all $\ell$ in $1, \dots, K$, 
$$ 1-\frac{K+1}{n-\ell} \geq 1-\frac{K+1}{n-K}.$$
Next,
\begin{align}
    &\sum_{ i_1< \dots < i_{K+1}}\sum_{  j_1< \dots < j_{K+1} } \pr [~\Delta_{ i_1, \dots, i_{K+1}}\cap\Delta_{j_1, \dots , j_{K+1}}~] \nonumber\\
    &= {n \choose {K+1}}{n-K-1 \choose {K+1}}\left(\frac{{1}}{{n-1 \choose K}}\right)^{2(K+1)}\nonumber\\
    &\qquad \cdot \left(\frac{{n-2K-3 \choose K}}{{n-1 \choose K}}\right)^{n-2(K+1)}\nonumber\\
     &\leq 
      {n \choose {K+1}}{n-K-1 \choose {K+1}}\left(\frac{{1}}{{n-1 \choose K}}\right)^{2(K+1)}\nonumber\\
    &\qquad \cdot \left(\frac{n-2K-3}{n-1}\right)^{K(n-2(K+1))}\label{eq:gen_2out-bound-doublesum-2}\\
    &=
      \frac{n!}{(n-2(k+1))!((K+1)!)^2}\left(\frac{{K!}}{{n-1 \choose K}}\right)^{2(K+1)}\nonumber\\
    &\qquad \cdot \left(\frac{n-2K-3}{n-1}\right)^{K(n-2(K+1))}\nonumber\\
     &= \frac{(K!)^{2(K+1)}}{((K+1)!)^2}
      \frac{n(n-1)\dots (n-2K-3)}{(n (n-1) \dots (n-K))^{2(K+1)}}\nonumber\\
    &\qquad \cdot \left(\frac{n-2K-3}{n-1}\right)^{K(n-2(K+1))}\nonumber\\
    &\leq \frac{(K!)^{2(K+1)}}{((K+1)!)^2} \cdot \frac{n^{2(K+1)}}{(n-K)^{2K(K+1)}}\nonumber\\
    &\qquad \cdot \left(1-\frac{2(K+1)}{n-1}\right)^{K(n-2(K+1))}
      \label{eq:gen_2out-bound-doublesum},
\end{align}{}
where (\ref{eq:gen_2out-bound-doublesum-2}) follows from (\ref{eq:ratio}).Substituting (\ref{eq:gen_2out-bound-singlesum}) and (\ref{eq:gen_2out-bound-doublesum}) in (\ref{eq:gen_2out-union-lb}), we obtain
\begin{align}
&\pr[~ Z_n \geq 1 ~] \nonumber\\
& \geq \sum_{  i_1<i_2\dots<i_{K+1}} \pr [~\Delta_{i_1\dotsi_{K+1}}~]\nonumber\\
&-\sum_{ i_1<i_2\dots<i_{K+1}}\sum_{ j_1<j_2\dots<j_{K+1}} \pr [~\Delta_{i_1\dotsi_{K+1}}\cap\Delta_{j_1,\dots,j_{K+1}}~]\nonumber\\
& \geq \frac{({K!})^K}{{K+1}}\cdot \frac{1}{n^{(K^2-1)}} \left(1-\frac{K+1}{n-K}\right)^{K(n-K-1)}\nonumber\\
&\qquad -\frac{(K!)^{2(K+1)}}{((K+1)!)^2} \cdot \frac{n^{2(K+1)}}{(n-K)^{2K(K+1)}}\nonumber\\
    &\qquad \cdot \left(1-\frac{2(K+1)}{n-1}\right)^{K(n-2(K+1))}\nonumber\\
& = \frac{({K!})^K e^{-K(K+1)}}{{K+1}}\cdot \frac{1}{n^{(K^2-1)}}  (1+\oo(1))  \label{eq:gen_upperrbound}\\
&= \Omega\left(\frac{1}{n^{K^2-1}} \right).\nonumber
\end{align}{}
\myendpf
In view of (\ref{eq:gen_p_con_upper_bound}), we then obtain for $K\geq2$ that
\[
\pr [~\hh \text{~is connected}~]  = 1-\Omega\left(\frac{1}{n^{K^2-1}} \right).
\]

\subsection{Bounding the sum in (\ref{eq:con_1_law_last_step})}
Recall that $K$ is a fixed positive integer $\geq 2$ and under the constraint (\ref{eq:OneLawConditions})
we  have $K+2 \leq \lfloor \frac{n}{2} \rfloor$,
and therefore the sum in (\ref{eq:con_1_law_last_step}) is  not empty. Let
\begin{equation}
\left(\frac{x}{n}\right)^{x(K-1)} = e^{(K-1) f_n (x) },
\quad x \geq 1
\label{eq:Exponentiation}
\end{equation}
with
\[
f_n (x) = x \log \left ( \frac{x}{n} \right )
        = x \left ( \log x - \log n \right ).
\]
Observe that 
$r \rightarrow f_n(r)$ decreases monotonically on the interval
$r=1, \ldots, \lfloor \frac{n}{e} \rfloor$
and increases monotonically thereafter on the interval
$r= \lfloor \frac{n}{e} \rfloor + 1, \ldots , \lfloor \frac{n}{2} \rfloor$. Therefore,
\begin{eqnarray}
\lefteqn{
\max \left (
f_n(r), \ r=K+2, \ldots , \left \lfloor \frac{n}{2} \right \rfloor
\right )
} & &
\nonumber \\
&=&
\max \left (
f_n(K+2),
f_n \left ( \left \lfloor \frac{n}{2} \right \rfloor \right )
\right ) .
\label{eq:IntermediaryBound}
\end{eqnarray}
From (\ref{eq:OneLawConditions}), we have $K+2 \leq \lfloor \frac{n}{e} \rfloor $. Next, we show that
\begin{equation}
f_n \left ( \left \lfloor \frac{n}{2} \right \rfloor \right )
\leq
f_n(K+2)
\label{eq:WhichIsLarger}
\end{equation}
for all $n$ large enough, say $n \geq n(K)$ 
for some finite integer $n(K)$ which depends on $K$.
(\ref{eq:WhichIsLarger}) is equivalent to
\[
\left \lfloor \frac{n}{2} \right \rfloor
\log
\left( \frac{ \left \lfloor \frac{n}{2} \right \rfloor }{n} \right )
\leq
(K+2) \left ( \log (K+2) - \log n \right ),
\]
a condition can be expressed as
\[
n \left ( 
\frac{ \left \lfloor \frac{n}{2} \right \rfloor }{n}
\right )
\log
\left( \frac{ \left \lfloor \frac{n}{2} \right \rfloor }{n} \right )
\leq
(K+2) \left ( \log (K+2) - \log n \right ).
\]
The mapping $t \rightarrow t \log t $ is monotone increasing
on the interval $( e^{-1}, \infty)$. Since $\left \lfloor \frac{n}{2} \right \rfloor \leq \frac{n}{2}$, for 
the inequality (\ref{eq:WhichIsLarger}) to hold, if suffices to show
\begin{equation}
- \left ( \frac{n}{2} \right ) \log 2
\leq
(K+2) \left ( \log (K+2) - \log n \right )
\label{eq:WhichIsLarger2}
\end{equation}
for all $n$ satisfying the constraint 
\[
\frac{1}{e} < \frac{1}{n} \left \lfloor \frac{n}{2} \right \rfloor  .
\]
It is easy to see that this occurs for all $n > 4$, which is in fact automatically guaranteed under (\ref{eq:OneLawConditions}).
Condition (\ref{eq:WhichIsLarger2}) can be simplified to yield
\begin{equation}
\log n
\leq
\left ( \frac{\log 2}{2(K+2)} \right ) \cdot n
+ \log (K+2).
\label{eq:WhichIsLarger3}
\end{equation}
It can be verified that 
(\ref{eq:WhichIsLarger3}) holds as an equality for $n=4(K+2)$
and a strict inequality for all $n > 4(K+2)$. Setting $n(K) = 4(K+2) $ is therefore sufficient for (\ref{eq:WhichIsLarger2})
(hence (\ref{eq:WhichIsLarger})) to hold.
Using (\ref{eq:Exponentiation}),
(\ref{eq:IntermediaryBound}) and (\ref{eq:WhichIsLarger})
we get
\begin{eqnarray}
\lefteqn{\max \left ( \left(\frac{r}{n}\right)^{r(K-1)} : \ r = K+2 ,
\ldots, \left \lfloor \frac{n}{2} \right \rfloor \right )} &&
\nonumber \\
&=&
\left(\frac{K+2}{n}\right)^{(K+2)(K-1)}    \hspace{ 1cm}
\nonumber
\end{eqnarray}
for all $n \geq n(K)$ yielding
\[
\sum_{r=K+2}^{ \lfloor \frac{n}{2} \rfloor }
\left(\frac{r}{n}\right)^{r(K-1)}
\leq
\left \lfloor \frac{n}{2} \right \rfloor
\cdot
\left(\frac{K+2}{n}\right)^{(K+2)(K-1)} .
\]
Substituting in (\ref{eq:con_1_law_last_step}) and noting that $P(n;K)=1-\bP{ C_n(K)^c }$, we  obtain (\ref{eq:LowerBoundForConnectivity}.
\myendpf
\end{document}